\documentclass[twoside]{ae100prg}
\usepackage{graphicx}
\usepackage[breaklinks]{hyperref}
\usepackage{booktabs}

\newcommand{\be}{\begin{equation}}
\newcommand{\ee}{\end{equation}}

\newcommand{\ba}{\begin{eqnarray}}
\newcommand{\ea}{\end{eqnarray}}

\newcommand\bfedit\bf

\newcommand{\Bf}{{magnetic field}}
\newcommand{\Bfs}{{magnetic fields}}
\newcommand{\Ef}{{electric  field}}
\newcommand{\Efs}{{electric fields}}

\newcommand{\NS}{neutron star}
\newcommand{\NSs}{{neutron stars}}
\newcommand{\EM}{electromagnetic}
\newcommand{\BH}{{black hole}}
\newcommand{\BHs}{{black holes}}
\newcommand{\Sc}{Schwarzschild}
\newcommand{\ms}{magnetosphere}
\newcommand{\mss}{magnetospheres}

\begin{document}
\title{Hair of astrophysical black holes}


\author{Maxim Lyutikov$^1$}

\address{$^1$ Department of Physics, Purdue University, 
 525 Northwestern Avenue,
West Lafayette, IN
47907-2036, USA
and
INAF - Osservatorio Astrofisico di Arcetri, 
Largo Enrico Fermi 5, I - 50125 Firenze,  Italia }

\email{lyutikov@purdue.edu}

\begin{abstract}
 The ``no hair''
  theorem is not   applicable to  black holes formed from collapse of a
  rotating neutron star.  Rotating neutron stars can self-produce particles via vacuum
  breakdown forming a highly conducting plasma magnetosphere
  such that magnetic field lines are effectively ``frozen-in'' the star both before and during collapse.
  In the limit of no resistivity, this introduces a topological constraint which prohibits the magnetic field
  from sliding off the newly-formed event horizon. As a result, during
  collapse of a neutron star into a black hole, the latter conserves the number of
  magnetic flux tubes $N_B = e \Phi_\infty /( \pi c \hbar)$, where $\Phi_\infty$
 is the
  initial magnetic flux through the hemispheres of the progenitor and
  out to infinity.   The
  black hole's magnetosphere  subsequently relaxes to the split monopole
  magnetic field geometry with self-generated currents outside the
  event horizon. The dissipation of the resulting equatorial current
  sheet leads to a slow loss of the anchored flux tubes, a process
  that balds the black hole on long resistive time scales rather than
  the short light-crossing time scales expected from the vacuum
  ``no-hair'' theorem.
\end{abstract}

\section{Introduction}

The ``no hair'' theorem \cite{MTW} postulates that all black hole
solutions of the Einstein-Maxwell equations of gravitation and
electromagnetism in general relativity can be completely characterized
by only three externally observable classical parameters: mass,
electric charge, and angular momentum.  The key point in the classical
proof \cite{1972PhRvD...5.2439P} is that the outside medium is a vacuum. In contrast, the
surroundings of astrophysical high energy sources like pulsars and
\BHs\ can rarely be treated as vacuum
\cite{GoldreichJulian,Blandford:1977,1992MNRAS.255...61M}.  The
ubiquitous presence of \Bfs\ combined with high (often relativistic)
velocities produce inductive \Efs\ with electric potential drops high
enough to break the vacuum via various radiative effects (curvature
emission followed by a single photon pair production in \Bf, or
inverse Compton scattering followed by a two photon pair production).
For example, in case of \NSs\ the rotation of the \Bf\ lines frozen
into the crust generates an inductive electric field, which, due to
the high conductivity of the \NS\ interior, induces surface
charges. The \Ef\ of these induced surface charges has a component
parallel to the dipolar \Bf.  These parallel \Efs\ accelerate charges
to the energy $ {\cal E} \sim e B_s R_s ( \Omega R_0/c)^2$, where
$B_s$ and $R_s$ are the surface \Bf, radius of a \NS\ and $\Omega$ is
the angular rotation frequency. The resulting primary beam of leptons
produces a dense secondary plasma via vacuum breakdown. Thus, in case
of \NSs\ the electric charges and currents are self-generated: no
external source is needed.  Rotating black holes can also lead to a
similar vacuum break-down \cite{Blandford:1977}.

We demonstrated that contrary to the prediction of the
``no hair'' theorem, the collapse of a rotating \NS\ into the \BH\
results in a formation of a long lived self-generated conducting BH
\ms.  This results from the violation of the key assumption of the
``no hair'' theorem, that the outside is vacuum, and allows a \BH\ to
preserve open magnetic flux tubes that initially connect to the \NS\
surface.

\section{The Black Hole Hair:  the Conserved Poloidal Magnetic Flux}

Consider collapse of a rotating \NS\ into the BH. 
Before the onset of the collapse, the electric currents within the
\NS\ create poloidal \Bf. Rotation of the poloidal \Bf\ lines and the
resulting inductive electric field lead to the creation, through
vacuum breakdown, of the conducting plasma and poloidal electric
currents.  The presence of a conducting plasma then imposes a
topological constraint, that the \Bf\ lines which initially were
connecting the \NS\ surface to the infinity must connect the \BH\
horizon to the infinity.

During the collapse, as the surface of a \NS\ approaches the horizon,
the closed \Bf\ lines will be quickly absorbed by the \BH, while the
open field lines (those connecting to infinity) have to remain open by
the frozen-in condition.  Thus, a \BH\ can have only open fields
lines, connecting its horizon to the infinity. There is a well known
solution that satisfies this condition: an exact split monopolar
solution for rotating \ms\ due to \cite{Michel73}; it was generalized
to \Sc\ metrics by \cite{Blandford:1977}. We recently found an exact
non-linear {\it time-dependent } split monopole-type structure of
\mss\ driven by spinning and collapsing \NS\ in \Sc\ geometry
\cite{2011PhRvD..83l4035L}. We demonstrated that the collapsing \NS\
enshrouded in a self-generated conducting \ms\ does not allow a quick
release of the \Bfs\ to infinity.

Thus, if a collapsing \BH\ can self-sustain the plasma production in
its \ms, the magnetic field lines that were initially connecting the
\NS\ surface to infinity will connect the \BH\ horizon to the
infinity.  Each hemisphere then keeps the magnetic flux that was
initially connected to the infinity.  For a \NS\ with the surface \Bf\
$B_{NS}$ and the initial pre-collapse radius $R_{NS}$ and period
$P_{\rm NS}$, the magnetic flux through each hemisphere connecting to
infinity is $\Phi_\infty \approx 2 \pi^2 B_{NS} R_{NS}^3 /(P_{\rm NS}
c)$ \cite{GoldreichJulian}. Using quantization of the magnetic flux
\cite{LLV}, this corresponds to a conserved quantum number of
magnetic flux tubes
\be
N_B =  e \Phi_\infty /( \pi c \hbar) = {2 \pi  B_{NS}  e R_{NS} ^3}/({c^2 {\hbar} P_{NS} }) = 10^{41} {B_{NS} \over 10^{12} {\rm G}} \, {P_{NS}  \over 1 {\rm msec}} .
\label{NB}
\ee
This quantum number is the \BH\ ``hair'': an observer at infinity can
measure the corresponding Poynting flux and infer the number $N_B$.

\section{Numerical Simulations}

We have performed numerical simulations that confirm the basic
principle that the ``no-hair'' theorem and related time-dependent
vacuum simulations are not applicable to a plasma-filled black hole
magnetosphere.  We do not model the process of vacuum breakdown and
the subsequent formation of a plasma-filled magnetosphere.  Instead,
we assume the \NS\ already created a plasma-filled magnetosphere (or
that the black hole self-generates a plasma-filled magnetosphere), and
we assume that the \NS\ has already collapsed to a black hole.
Only once an event horizon has formed would the magnetic
field begin to slip-off the black hole in vacuum, so starting with
an event horizon should be a strong enough test -- one should not
have to follow the collapse of the \NS\ to a black hole as long as a
plasma is present.  The goal of the simulations is to measure the
decay timescale of the magnetic flux threading the event
horizon of the black hole: $\Phi_{\rm EM} = (1/2)\int_S dS |B^r|$ as
integrated over the surface ($S$) of the black hole horizon.  We show
that the magnetic dipole decay seen in vacuum solutions
is avoided or delayed by three effects:
1) presence of plasma and self-generation of toroidal currents ;
2) black hole spin induced poloidal currents ;
and 3) plasma pressure support of current layers generated internally by dissipating currents.
These effects cause the field to avoid vacuum-like decay of the dipole
magnetic field and help support the newly-formed split-monopole \Bf\
against magnetic reconnection.

These GRMHD simulations use the fully conservative, shock-capturing
GRMHD scheme called HARM \cite{2006MNRAS.368.1561M}
using Kerr-Schild coordinates in the Kerr metric for a sequence of
spins. 

We perform simulations that either use the force-free or use the fully
energy-conserving MHD equations of motions.  These approximate,
respectively, the limits of radiatively efficient emission and
radiatively inefficient emission once the plasma has been generated.
That is, if the electromagnetic field dominates the rest-mass and
internal energy density over most of the volume outside current
sheets, then the force-free limit corresponds to an instantaneous loss
(such as radiation) of magnetic energy dissipated in current sheets,
while the fully energy-conserving MHD limit without cooling
corresponds to all dissipated energy going into internal+kinetic
energy that remains in the system and sustains the current sheet
against dissipation.  A non-energy-conserving system of equations or
simulation code would be unable to properly follow the energy
conservation process of electromagnetic dissipation within the current
sheet that leads to plasma formation there. The force-free
electrodynamics equations of motion are not solely relied upon because
they are undefined within current sheets and any particular resistive
force-free electrodynamics equations
\cite{LyutikovTear,gruz08,2011arXiv1107.0979L} still leave some
degree of ambiguity in how the resistivity would map onto the full
magnetohydrodynamical (MHD) equations.  For the MHD equations, an
ideal $\gamma=4/3$ gas equation of state is chosen, which can be
considered as mimicking a radiatively inefficient high-energy particle
distribution component generated by the dissipation of the currents
within the reconnecting layer.

\begin{figure}[htb]
\includegraphics[width=0.99\linewidth]{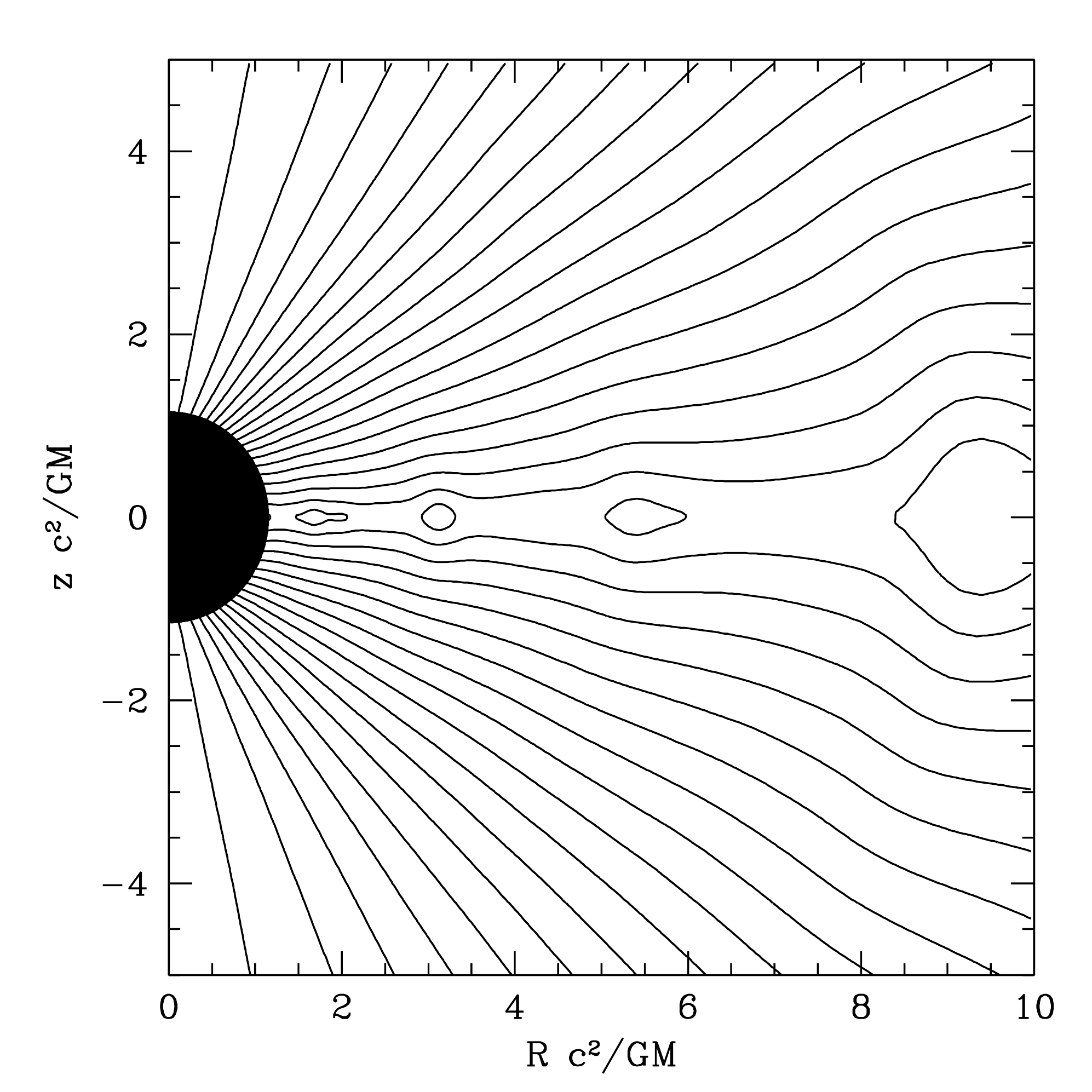}

\caption{ A contour plot of the magnetic flux ($\Psi=R A_\phi$)
  showing the inner (cylindrical radius) $R<10GM/c^2$ for the MHD
  $a=0.99$ model described in the text.  The structure of the \ms\ relaxes to monopolar-like
  solution, as predicted by \cite{2011PhRvD..83l4035L}. Note also the
  development of the tearing modes and the formation of magnetic
  islands in the equatorial current sheet.}
\label{aphi}
\end{figure}

\section{Astrophysical applications}

The fact that  isolated \BHs\  formed in a collapse of rotating \NSs\ can retain their  open magnetic flux for times much longer than the collapse time  implies that isolated BHs can spindown electromagnetically, converting the rotational energy in the \EM\ wind. This can be important for short GBRs: in \cite{}  we address the key problem of the \NS\ merger paradigm in application to short GRBs, the presence of energetic prompt tails and flares at very long time scales, orders of magnitude longer than the active stage of the merger.  We identify the prompt GRB spike as coming from the energy  dissipation of the wind powered by a transient accretion torus surrounding the newly formed GRB. It's duration is limited by the life of the torus, tens to hundreds of milliseconds. The long extended emission comes from wind powered by the isolated rotating BH, that produces equatorially-collimated outflow.
It's duration is limited by the retention time scale of the \Bf, and it contains {\it more total energy than the prompt spike}.

\begin{figure}[h!]
\includegraphics[width=0.99\linewidth]{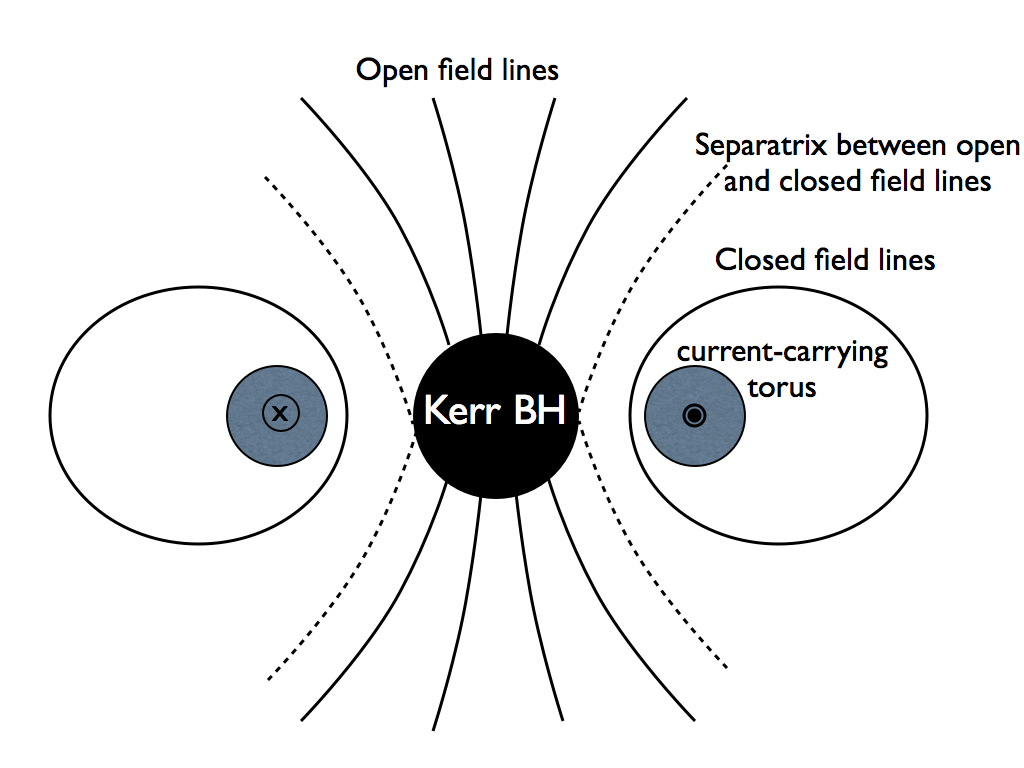}
\caption{Schematic presentations of magnetic flux surfaces in the BH-torus system. Toroidal electric current in the torus creates poloidal \Bf. The field lines that intersect the BH are twisted by the rotation of the space-time (carry poloidal electric current) and open up to infinity.  There are two types of \Bf\ lines separated by a separatrix (dashed lines): closed field lines and open  \Bf\ lines that intersect the Kerr BH. (The section shows only the poloidal component of the \Bf.) After the torus is accreted,  the open \Bf\ lines remain on the BH, relaxing to a twisted  monopolar structure \cite{Blandford:1977,Michel73}. }
\label{torus-pict}
\end{figure}

Thus, the proposed model for short GRBs implies a different type of collimation of the outflow  than the conventionally envisioned jet-like structure, at least in the prompt tail stage.
 An observer on the axis see only the axially collimated  prompt emission generated by the BH-torus system, while an observer at medium polar angles sees both the prompt spike and the equatorially-collimated extended tail. 

In addition,  the efficiency of energy extraction of the \BH\ spin energy  during  episodic accretion of magnetized blobs can exceed the average  mass accretion rate $\dot{M} c^2$,  while the total extracted energy can exceed the accreted rest mass. This phenomenon   can lead to production of powerful flares via accretion of fairly small amount of matter.
\section*{References}

 \bibliography{/Users/maxim/Home/Research/BibTex}

\end{document}